\documentclass[aps,prl,showpacs]{revtex4}
\usepackage{graphicx}
\usepackage{amsfonts,amssymb,amsmath}
\usepackage{bm}

\begin{document}

\title{Differential Realizations of the Two-Mode Bosonic and Fermionic
Hamiltonians: A unified Approach }
\date{\today}
 
\author{Hayriye T\"{u}t\"{u}nc\"{u}ler}
\email{tutunculer@gantep.edu.tr}
\affiliation{Department of Physics, Faculty of Engineering 
University of Gaziantep,  27310 Gaziantep, Turkey}
\author{Ramazan Ko\c{c}}
\email{koc@gantep.edu.tr}
\affiliation{Department of Physics, Faculty of Engineering 
University of Gaziantep,  27310 Gaziantep, Turkey}

\begin{abstract}
A method is developed to determine the eigenvalues and
eigenfunction of two-boson $2\times 2$ matrix Hamiltonians include
a wide class of quantum optical models. The quantum Hamiltonians
have been transformed in the form of the one variable differential
equation and the conditions for its solvability have been
discussed. We present two different transformation procedure and
we show our approach unify various approaches based on Lie
algebraic technique. As an application, solutions of the
modified Jaynes-Cummings and two-level Jahn Teller Hamiltonians
are studied.
\end{abstract}

\maketitle

\section{Introduction}

The algebraic techniques have been proven to be useful in the description of
the physical problems in a variety of fields. Recently a new algebraic
approach, essentially improving both analytical and numerical solution of
the problems, has been suggested and developed for some nonlinear quantum
optical systems\cite{klim2,kara4,gab,alv}. Most of such developments are
mainly based on linear Lie algebras, but it is evident that there is no
physical reason for symmetries to be only linear. Nonlinear Lie algebra
techniques and their relations to the nonlinear quantum optical systems have
been discussed\cite{beck,brihaye,tjin,abd,sunil}. In both cases finite part
of spectrum of the corresponding Hamiltonian can be exactly obtained in
closed forms and these systems are known as quasi-exactly-solvable(QES)
termed by Turbiner and Ushveridze\cite{turb1}. It has been proven that the
single boson Hamiltonians also lead to a QES under some certain constraints%
\cite{dolya1,dolya2}.

In recent years there has been a great deal of interest in quantum optical
models which reveal new physical phenomena described by the Hamiltonians
expressed as nonlinear functions of Lie algebra generators or boson and/or
fermion operators\cite{elber,kara1,kara2,delgado,klim}. Such systems have
often been analyzed by using numerical methods, because the implementation
of the Lie algebraic techniques to solve those problems is not very
efficient and most of the other analytical techniques do not yield simple
analytical expressions. They require tedious calculations\cite%
{kara3,band,jurco}.

The aim of this paper is to determine solvability of the two-boson $2\times
2 $ matrix Hamiltonians and discuss their applications and possible symmetry
groups. The results of our procedure include the solutions of the
Hamiltonians that possessing $su(2)$, $su(1,1)$, $Sp(4,R)$, $osp(2,1)$ and $%
osp(2,2)$ symmetries. The procedure presented in this paper also leads to
the constructions of the nonlinear Lie algebras\cite{kumar}. The
Hamiltonians discussed here are not only mathematically interesting but they
have potential interest in physics\cite{per,baj,qu}. In order to keep our
discussion simple we concentrate our attention to the solution of the
Hamiltonians that include two boson and one fermion. Therefore we provide a
first step toward the extension of the technique to the solution of the
multi boson fermion systems.

The paper is organized as follows: In section 2, we construct a Hamiltonian
including two boson operators in an arbitrary order and one fermion
operator. Solutions of the Hamiltonians by using the invariance of the
number operator have been discussed. As a practical example, solution of the
Modified Jaynes-Cummings and two level Jahn Teller Hamiltonians have been
obtained. In section 3 we present two transformation procedure that is
appropriate to determine the conditions of the (quasi)exact solvability of
the Hamiltonian. In section 4 we discuss the symmetry properties of the
Hamiltonian. The results of our discussion are that the Hamiltonian contains
various Lie (super)algebras, namely, $su(2),su(1,1),Sp(4,R),osp(2,1)$ and $%
osp(2,2)$. We also point out that nonlinear Lie algebras can be constructed
as spectrum generating algebras. As an application we visit Jaynes-Cummings
Hamiltonian including Kerr nonlinearity. The importance of our approach is
that it provides a unification of the various approaches. Finally, \ in
section 5 we comment on the validity of our method and suggest the possible
extensions of the problem.

\section{Two boson one fermion Hamiltonian and its differential realization}

Two mode bosonic $2\times 2$ matrix Hamiltonians play an important role in
nonlinear quantum optical systems. The Hamiltonians of such systems can be
generalized as follows:

\begin{eqnarray}
H &=&\sum_{k_{i}}\alpha _{k_{1,}k_{2},k_{3},k_{4}}B^{k}+\sigma
_{0}\sum_{\ell _{i}}\beta _{\ell _{1},\ell _{2},\ell _{3},\ell _{4}}B^{\ell
}+  \nonumber \\
&&\sigma _{+}\sum_{m_{i}}\gamma _{m_{1,}m_{2},m_{3},m_{4}}B^{m}+\sigma
_{-}\sum_{n_{i}}\delta _{n_{1,}n_{2},n_{3},n_{4}}B^{n}.  \label{1}
\end{eqnarray}%
The bosonic operator $B^{\upsilon }$ reads as follows:%
\begin{equation}
B^{\upsilon }=(a_{1}^{+})^{\upsilon _{1}}(a_{1})^{\upsilon
_{2}}(a_{2}^{+})^{\upsilon _{3}}(a_{2})^{\upsilon _{4}}  \label{1a}
\end{equation}%
The constants $\alpha _{i},\beta _{i},\gamma _{i}$ and $\delta _{i}$ are
related to the parameters of the physical Hamiltonian and $k_{i},\ell
_{i},m_{i}$ and $n_{i}$ determine order of the interaction. The boson
creation, $a_{1}$, $a_{2}$, and annihilation $a_{1}^{+}$, $a_{2}^{+}$
operators obey the usual commutation relations

\begin{equation}
\left[ a_{i},a_{i}\right] =\left[ a_{i}^{+},a_{i}^{+}\right] =0,\quad \left[
a_{i},a_{j}^{+}\right] =\left\{ 
\begin{array}{c}
1,i=j \\ 
0,i\neq j%
\end{array}%
\right.  \label{2}
\end{equation}%
and $\sigma _{\pm ,0}$ are Pauli matrices:%
\begin{equation}
\sigma _{-}=\left( 
\begin{array}{cc}
0 & 0 \\ 
1 & 0%
\end{array}%
\right) ,\quad \sigma _{+}=\left( 
\begin{array}{cc}
0 & 1 \\ 
0 & 0%
\end{array}%
\right) ,\quad \sigma _{0}=\left( 
\begin{array}{cc}
1 & 0 \\ 
0 & -1%
\end{array}%
\right) .  \label{3}
\end{equation}%
The number operator of such system can be expressed in the form%
\begin{equation}
N=sa_{1}^{+}a_{1}+pa_{2}^{+}a_{2}+r\sigma _{0}  \label{4}
\end{equation}%
and satisfy the commutation relations%
\begin{eqnarray}
\lbrack N,a_{1}^{+}] &=&sa_{1}^{+},\quad \lbrack N,a_{1}]=-sa_{1},\quad
\lbrack N,a_{2}^{+}]=pa_{2}^{+},\quad \lbrack N,a_{2}]=-pa_{2}\quad 
\nonumber \\
\lbrack N,\sigma _{\pm }] &=&\pm 2r\sigma _{\pm }\quad \lbrack N,\sigma
_{0}]=0.  \label{5}
\end{eqnarray}%
The action of the number operator on the state $\left|
n_{1},n_{2}\right\rangle $ is given by%
\begin{equation}
N\left| n_{1},n_{2}\right\rangle =(sn_{1}+pn_{2}+r)\left|
n_{1},n_{2}\right\rangle  \label{6}
\end{equation}%
The eigenvalue equation (\ref{5}) leads to the following solution:%
\begin{equation}
\left| n_{1},n_{2}\right\rangle =(a_{1}^{+})^{j}\phi _{1}\left(
a_{2}^{+}(a_{1}^{+})^{-\frac{p}{s}}\right) \left| \uparrow \right\rangle
+(a_{1}^{+})^{j+\frac{2r}{p}}\phi _{2}\left( a_{2}^{+}(a_{1}^{+})^{-\frac{p}{%
s}}\right) \left| \downarrow \right\rangle .  \label{7}
\end{equation}%
where $\left| \uparrow \right\rangle $ and $\left| \downarrow \right\rangle $
are up and down states and $j$ is given by%
\begin{equation}
j=n_{1}+\frac{p}{s}n_{2}.  \label{8}
\end{equation}%
If $N$ and $H$ commute, the function (\ref{7}) is also eigenfunction of the $%
H$. Therefore it is worth to seek the conditions for the commutation of $H$
and $N$. This can easily be done by using the commutation relations (\ref{2}%
) and (\ref{5}) and we obtain the following relation%
\begin{eqnarray}
\lbrack N,H] &=&\sum_{k_{i}}[s(k_{1}-k_{2})+p(k_{3}-k_{4})]\alpha
_{k_{1}k_{2}k_{3}k_{4}}B^{k}+  \nonumber \\
&&\sigma _{0}\sum_{\ell _{i}}[s(\ell _{1}-\ell _{2})+p(\ell _{3}-\ell
_{4})]\beta _{\ell _{1}\ell _{2}\ell _{3}\ell _{4}}B^{\ell }+  \nonumber \\
&&\sigma _{+}\sum_{m_{i}}[s(m_{1}-m_{2})+p(m_{3}-m_{4})+r]\gamma
_{m_{1}m_{2}m_{3}m_{4}}B^{m}+  \label{9} \\
&&\sigma _{-}\sum_{n_{i}}[s(n_{1}-n_{2})+p(n_{3}-n_{4})-r]\delta
_{n_{1}n_{2}n_{3}n_{4}}B^{n}  \nonumber
\end{eqnarray}%
Then the constant of motion $N$ and $H$ commute when the fallowing set of
equation is satisfied%
\begin{eqnarray}
s(k_{1}-k_{2})+p(k_{3}-k_{4}) &=&0  \nonumber \\
s(\ell _{1}-\ell _{2})+p(\ell _{3}-\ell _{4}) &=&0  \nonumber \\
s(m_{1}-m_{2})+p(m_{3}-m_{4})+2r &=&0  \label{10} \\
s(n_{1}-n_{2})+p(n_{3}-n_{4})-2r &=&0.  \nonumber
\end{eqnarray}

Now we demonstrate application of the procedure on a physical example. We
study modified Jaynes-Cummings Hamiltonian and this give us an opportunity
to test our approach because those Hamiltonians have been studied in
literature.

\subsection{The modified Jaynes-Cummings Hamiltonian}

The modified Jaynes-Cummings Hamiltonian have been constructed to
investigate single two level atom placed in the common domain of two
cavities interacting with two quantized modes. It is given by\cite{bo}: 
\begin{equation}
H=\omega (a_{1}^{+}a_{1}+a_{2}^{+}a_{2})+\frac{\omega _{0}}{2}\sigma
_{0}+\lambda _{1}(a_{1}\sigma _{+}+a_{1}^{+}\sigma _{-})+\lambda
_{2}(a_{2}\sigma _{+}+a_{2}^{+}\sigma _{-}).  \label{11}
\end{equation}%
When the parameters%
\begin{equation}
\alpha _{i,j,k,l}=\beta _{i,j,k,l}=\gamma _{i,j,k,l}=\delta _{i,j,k,l}=0
\label{12}
\end{equation}%
except that%
\begin{eqnarray}
\alpha _{1,1,0,0} &=&\alpha _{0,0,1,1}=\omega ,\beta _{0,0,0,0}=\frac{\omega
_{0}}{2},  \nonumber \\
\gamma _{0,1,0,0} &=&\delta _{1,0,0,0}=\lambda _{1},\gamma _{0,0,0,1}=\delta
_{0,0,1,0}=\lambda _{2}  \label{13}
\end{eqnarray}%
then the Hamiltonians (\ref{1}) and (\ref{11}) are identical. The condition (%
\ref{10}) is satisfied when $s=p=2r$. Before going further in the following
we use the Bargmann-Fock representation, where creation and annihilation
operators are replaced by multiplication and differentiation operators:

\begin{equation}
a_{i}^{+}=z_{i},\quad a_{i}=\frac{d}{dz_{i}}  \label{14}
\end{equation}%
with respect to complex variable $z_{i}$. The eigenfunction (\ref{7}) is of
the form

\begin{equation}
\left| n_{1},n_{2}\right\rangle =(z_{1})^{j}\phi \left( x\right) \left|
\uparrow \right\rangle +(z_{1})^{j+1}\phi \left( x\right) \left| \downarrow
\right\rangle  \label{15}
\end{equation}%
where $x=(z_{1})^{-1}z_{2}$. The solution of this system describes a quantum
mechanical state of $H$ provided that $\phi (x)$ belong to the Bargmann-Fock
space. The scalar product should be complete and normalizable,

\begin{equation}
\int \overline{\phi (x)}\phi (x)e^{\int W(x)dx}d({Re}x)d({Im}x)<\infty
\label{16}
\end{equation}%
where $W(x)$ is the weight function. The eigenvalue equation of the modified
Jaynes-Cummings Hamiltonian can be written as%
\begin{equation}
H\left| n_{1},n_{2}\right\rangle =E\left| n_{1},n_{2}\right\rangle
\label{17}
\end{equation}%
Insertion of (\ref{14}) and (\ref{15}) into (\ref{17}) yield the following
two set of differential equation

\begin{eqnarray}
\left[ j\omega +\frac{\omega _{0}}{2}-E\right] \phi _{1}(x)+(j+1)\lambda
_{2}\phi _{2}(x)+(\lambda _{2}-x\lambda _{1})\frac{d\phi _{2}(x)}{dx} &=&0 
\nonumber \\
\left[ (j+1)\omega -\frac{\omega _{0}}{2}-E\right] \phi _{2}(x)+(\lambda
_{1}+x\lambda _{2})\phi _{1}(x) &=&0  \label{18}
\end{eqnarray}%
Bargmann-Fock space solution of the (\ref{18}) can easily be obtained and
they are given by%
\begin{eqnarray}
\phi _{1}(x) &=&C_{0}\left( \lambda _{2}-x\lambda _{1}\right) ^{j-n+1}\left(
\lambda _{1}+x\lambda _{2}\right) ^{n-1}  \nonumber \\
\phi _{2}(x) &=&C_{1}\left( \lambda _{2}-x\lambda _{1}\right) ^{j-n+1}\left(
\lambda _{1}+x\lambda _{2}\right) ^{n}  \label{19}
\end{eqnarray}%
where $n$ is an integer and eigenvalues of the Hamiltonian is given by%
\begin{equation}
E=\frac{1}{2}\left( (2j+1)\omega \pm \sqrt{4n(\lambda _{1}^{2}+\lambda
_{2}^{2})+(\omega _{0}-\omega )^{2}}\right) .  \label{20}
\end{equation}%
Consequently we have obtained exact result for the eigenvalues of the
modified Jaynes-Cummings Hamiltonian. The same result have been obtained in%
\cite{bo}, in the framework of the $su(2)$ algebra. The procedure given here
can be applied to obtain eigenfunction and eigenvalues of the various
physical Hamiltonians. In the following example we consider the solution of
the Jahn-Teller distortion problem.

\subsection{Two-Level Jahn-Teller Distortion Problem}

The well-known form of the JT Hamiltonian describing a two-level fermionic
subsystem coupled to two boson modes has been given by \cite{reik}:%
\begin{equation}
H=a_{1}^{+}a_{1}+a_{2}^{+}a_{2}+1+(\frac{1}{2}+2\mu )\sigma _{0}+2\kappa
\lbrack (a_{1}+a_{2}^{+})\sigma _{+}+(a_{1}^{+}+a_{2})\sigma _{-}].
\label{ed1}
\end{equation}%
Our task is now to demonstrate the Hamiltonian (\ref{ed1}) can be solved in
the framework of the procedure given previous section. It will be shown that
our approach relatively very simple when compared previous approaches. The
Hamiltonians (\ref{ed1}) and (\ref{1}) are identical, when the parameters
are constrained to:%
\begin{eqnarray}
\alpha _{1,1,0,0} &=&\alpha _{0,0,1,1}=\alpha _{0,0,0,0}=1,\quad \beta
_{0,0,0,0}=(\frac{1}{2}+2\mu ),  \nonumber \\
\gamma _{0,1,0,0} &=&\delta _{1,0,0,0}=\gamma _{0,0,1,0}=\delta
_{0,0,0,1}=2\kappa  \label{ed2}
\end{eqnarray}%
otherwise

\begin{equation}
\alpha _{i,j,k,l}=\beta _{i,j,k,l}=\gamma _{i,j,k,l}=\delta _{i,j,k,l}=0
\label{ed3}
\end{equation}%
The condition (\ref{10}) is satisfied when $s=-p=2r.$ In the Bargmann-Fock
space the eigenfunction (\ref{7}) takes the form%
\begin{equation}
\left| n_{1},n_{2}\right\rangle =(z_{1})^{j}\phi _{1}\left( x\right) \left|
\uparrow \right\rangle +(z_{1})^{j-1}\phi _{2}\left( x\right) \left|
\downarrow \right\rangle .  \label{ed4}
\end{equation}%
where $x=z_{1}z_{2}$ and $j=n_{1}-n_{2}.$ Since number operator $N$ and
Hamiltonian (\ref{ed1}) commute, they have the same eigenfunction.
Substituting (\ref{ed4}) into (\ref{ed1}) we obtain the following set of
equation:

\begin{eqnarray}
x\phi _{1}^{\prime }\left( x\right) +\kappa x\phi _{2}^{\prime }\left(
x\right) +\frac{1}{4}(3-2E+2j+4\mu )\phi _{1}\left( x\right) +\kappa
(1+j+x)\phi _{2}\left( x\right) &=&0  \label{ed50} \\
\kappa \phi _{1}^{\prime }\left( x\right) +x\phi _{2}^{\prime }\left(
x\right) +\frac{1}{4}(3-2E+2j-4\mu )\phi _{2}\left( x\right) +\kappa \phi
_{1}\left( x\right) &=&0  \label{ed5}
\end{eqnarray}%
These coupled differential equations represent the Schr\"{o}dinger equation
of the $E\otimes \epsilon $ Jahn-Teller system in the Bargmann's Hilbert
space. The equation is quasi-exactly-solvable and success of our analysis
leads to the solution of the various quantum optical systems. The physical
systems described by the differential equations (\ref{ed50}, \ref{ed5}) are
discussed in\cite{koc}.

The validity of the procedure depends on the choice of the $\alpha _{i}$, $%
\beta _{i}$, $\gamma _{i}$ and $\delta _{i}.$One can easily be obtain
various physical Hamiltonians by appropriate choice of $\alpha _{i}$, $\beta
_{i}$, $\gamma _{i}$ and $\delta _{i}$ and by considering the conditions
given in (\ref{10}).

\section{Transformation of the operators}

In this section we discuss transformation of the Pauli matrices and boson
operators. These transformations plays a key role to construct
(quasi)exactly solvable $2\times 2$ matrix Hamiltonians. The transformation
can be done by introducing the following similarity transformation induced
by the metric 
\begin{equation}
S=(a_{2}^{+})^{ca_{1}^{+}a_{1}+d\sigma _{+}\sigma _{-}}  \label{21}
\end{equation}%
where $c$ and $d$ are constants. Since $a_{1}$, and $a_{2}$ commute and $%
\sigma _{\pm ,0}$ also commute with the bosonic operators, the
transformation of $a_{1}$ and $a_{1}^{+}$ under $S$ can be obtained by
writing $a_{2}^{+}=e^{b}$, with $[a_{1},b]=[a_{1}^{+},b]=0$,%
\begin{eqnarray}
Sa_{1}S^{-1} &=&a_{1}(a_{2}^{+})^{-c}  \nonumber \\
Sa_{1}^{+}S^{-1} &=&a_{1}^{+}(a_{2}^{+})^{c}  \label{22}
\end{eqnarray}%
the transformation of $a_{2}$ and $a_{2}^{+}$ is as follows%
\begin{eqnarray}
Sa_{2}S^{-1} &=&a_{2}-(ca_{1}^{+}a_{1}+d\sigma _{+}\sigma
_{-})(a_{2}^{+})^{-1}  \nonumber \\
Sa_{2}^{+}S^{-1} &=&a_{2}^{+}  \label{23}
\end{eqnarray}%
and the transformations of the $\sigma _{\pm }$ are given by%
\begin{equation}
S\sigma _{\pm }S^{-1}=\sigma _{\pm }(a_{2}^{+})^{\pm d}.  \label{24}
\end{equation}

Before constructing one variable (quasi)exactly solvable differential
equation of the Hamiltonian (\ref{1}) under the transformations of the
bosonic and fermionic operators by $S$, let us consider the other
transformation operator:%
\begin{equation}
T=(a_{2})^{\varepsilon a_{1}^{+}a_{1}+\eta \sigma _{+}\sigma _{-}}
\label{25}
\end{equation}%
where $\varepsilon $ and $\eta $ are constants. By using the similar
arguments given in the previous transformation operations one can easily
obtain the following transformations:%
\begin{eqnarray}
Ta_{1}T^{-1} &=&a_{1}(a_{2}^{+})^{-\varepsilon }  \nonumber \\
Ta_{1}^{+}T^{-1} &=&a_{1}^{+}(a_{2})^{\varepsilon }  \nonumber \\
Ta_{2}T^{-1} &=&a_{2}  \label{26} \\
Ta_{2}^{+}T^{-1} &=&a_{2}^{+}+(\varepsilon a_{1}^{+}a_{1}+\eta \sigma
_{+}\sigma _{-})(a_{2})^{-1}  \nonumber \\
T\sigma _{\pm }T^{-1} &=&\sigma _{\pm }(a_{2}^{+})^{\pm \eta }.  \nonumber
\end{eqnarray}%
These transformations leads to the construction of the various differential
realizations of (\ref{1}), depending on the choice of $c,n,\varepsilon $ and 
$\eta .$ In the following we discuss the possible forms of the Hamiltonian (%
\ref{1}) and their solutions.

\section{S-Transformed Hamiltonian}

In this section we discuss the solvability of the Hamiltonian(\ref{1}). The
number operator $N$ describe the states of the corresponding Hamiltonian.
The transformation of $N$ under the operator $S$ is given by%
\begin{equation}
N^{\prime }=SNS^{-1}=(s-pc)a_{1}^{+}a_{1}+pa_{2}^{+}a_{2}+(2r-pd)\sigma
_{+}\sigma _{-}-r  \label{27}
\end{equation}%
The Hamiltonian (\ref{1}) is characterized by the total number of $a_{1}$
and $a_{2}$ bosons when the number operator $N$ commute with the whole
Hamiltonian. In the transformed case it is only the number of $a_{2}$ bosons
that characterize the system under the condition $c=s/p$ and $d=2r/p$. When
the representation is characterized by a fixed number $a_{2}^{+}a_{2}=j$,
then the transformed form the Hamiltonian can be expressed as one boson
operator $a_{1}$, under the condition (\ref{10}). The transformed form of
the Hamiltonian (\ref{1}) can be written as:%
\begin{eqnarray}
\widetilde{H} &=&SHS^{-1}=\sum_{k_{i}}\alpha
_{k_{1,}k_{2,}k_{4}}(a_{1}^{+})^{k_{1}}(a_{1})^{k_{2}}(j-\frac{s}{p}%
a_{1}^{+}a_{1}-\frac{2r}{p}\sigma _{+}\sigma _{-})^{k_{4}}  \nonumber \\
&&\sigma _{0}\sum_{\ell _{i}}\beta _{\ell _{1},\ell _{2},\ell
_{4}}(a_{1}^{+})^{\ell _{1}}(a_{1})^{\ell _{2}}(j-\frac{s}{p}a_{1}^{+}a_{1}-%
\frac{2r}{p}\sigma _{+}\sigma _{-})^{\ell _{4}}+  \label{28} \\
&&\sigma _{+}\sum_{m_{i}}\gamma
_{m_{1,}m_{2},m_{4}}(a_{1}^{+})^{m_{1}}(a_{1})^{m_{2}}(j-\frac{s}{p}%
a_{1}^{+}a_{1}-\frac{2r}{p}\sigma _{+}\sigma _{-})^{m_{4}}+  \nonumber \\
&&\sigma _{-}\sum_{n_{i}}\delta
_{n_{1,}n_{2},n_{4}}(a_{1}^{+})^{n_{1}}(a_{1})^{n_{2}}(j-\frac{s}{p}%
a_{1}^{+}a_{1}-\frac{2r}{p}\sigma _{+}\sigma _{-})^{n_{4}}.  \nonumber
\end{eqnarray}%
Note that $k_{3},\ell _{3},m_{3}$ and $n_{3}$ are eliminated by using the
conditions given in(\ref{10}) The difference between (\ref{1}) and (\ref{28}%
) is that while in the first the total number of $a_{1}$ and $a_{2}$ bosons
characterize the the system, in the later it is only the number of $a_{2}$
bosons that characterize the system. Therefore the representation is
characterized by a fixed number $j$ and in (\ref{28}), the Hamiltonian is
expressed in terms of one boson operator $a_{1}$. The transformed
Hamiltonian $\widetilde{H}$, in the Bargmann-Fock space, which play an
important role in the quasi-exact solution of the equation (\ref{1}). It can
be transformed in the form of the one dimensional differential equations in
the Bargmann-Fock space when the boson operators are realized as%
\begin{equation}
a_{1}=\frac{d}{dx},\quad a_{1}^{+}=x.  \label{29}
\end{equation}%
The basis function of the primed generators of the system is two component
spinor;

\begin{equation}
P_{n,m}(x)=\left( 
\begin{array}{c}
x^{0},x^{1},\cdots ,x^{n} \\ 
x^{0},x^{1},\cdots ,x^{m}%
\end{array}%
\right) .  \label{30}
\end{equation}%
Action of (\ref{28}) on the (\ref{30}), in the Bargmann-Fock space can be
written as:%
\begin{equation}
\widetilde{H}P_{n}(x)=\sum P_{n,m}(E)(x^{n},x^{m})  \label{31}
\end{equation}%
The wavefunction is itself the generating function of the energy
polynomials. The eigenvalues are then produced by the roots of such
polynomials. If the $E_{n,m}$ is a root of the polynomial $P_{n+1,m+1}(E)$,
the series (\ref{31}) terminates and $E_{n,m}$ belongs to the spectrum of
the corresponding Hamiltonian. The eigenvalues are then obtained by finding
the roots of such polynomials.

\section{T-Transformed Hamiltonian}

The constant of motion $N$ characterize the system can be transformed, by
the operator $T$ is given by%
\begin{equation}
N^{\prime }=TNT^{-1}=(s+p\varepsilon )a_{1}^{+}a_{1}+pa_{2}^{+}a_{2}+p\eta
\sigma _{+}\sigma _{-}+r.  \label{32}
\end{equation}%
The Hamiltonian (\ref{1}) can be characterized, in the transformed case, $%
\varepsilon =-s/p$ and $\eta =-2r/p$. Thus according to (\ref{32}) the
representation is characterized by a fixed number $a_{2}^{+}a_{2}=j-1$.
Therefore the transformed Hamiltonian includes one boson operator $a_{1}$,
when the condition (\ref{10}) is taken into consideration, it can be written
as:%
\begin{eqnarray}
H^{\prime } &=&THT^{-1}=\sum_{k_{i}}\alpha
_{k_{1,}k_{2,}k_{4}}(a_{1}^{+})^{k_{1}}(a_{1})^{k_{2}}(j-\frac{k_{4}}{k_{3}}-%
\frac{s}{p}a_{1}^{+}a_{1}-\frac{2r}{p}\sigma _{+}\sigma _{-})^{k_{4}} 
\nonumber \\
&&\sigma _{0}\sum_{\ell _{i}}\beta _{\ell _{1},\ell _{2},\ell
_{4}}(a_{1}^{+})^{\ell _{1}}(a_{1})^{\ell _{2}}(j-\frac{\ell _{4}}{\ell _{3}}%
-\frac{s}{p}a_{1}^{+}a_{1}-\frac{2r}{p}\sigma _{+}\sigma _{-})^{\ell _{4}}+
\label{33} \\
&&\sigma _{+}\sum_{m_{i}}\gamma
_{m_{1,}m_{2},m_{4}}(a_{1}^{+})^{m_{1}}(a_{1})^{m_{2}}(j-\frac{m_{4}}{m_{3}}-%
\frac{s}{p}a_{1}^{+}a_{1}-\frac{2r}{p}\sigma _{+}\sigma _{-})^{m_{4}}+ 
\nonumber \\
&&\sigma _{-}\sum_{n_{i}}\delta
_{n_{1,}n_{2},n_{4}}(a_{1}^{+})^{n_{1}}(a_{1})^{n_{2}}(j-\frac{n_{4}}{n_{3}}-%
\frac{s}{p}a_{1}^{+}a_{1}-\frac{2r}{p}\sigma _{+}\sigma _{-})^{n_{4}}. 
\nonumber
\end{eqnarray}%
The Hamiltonian can be expressed as one dimensional differential equation in
the Bargmann-Fock space.

In order to obtain exactly or QES Hamiltonians one can use the same basis
given in (\ref{30}). The physical Hamiltonians can be obtained and solved by
the choice of the appropriate values of the $s$ and $p$. Consequently we
have obtained two classes of Hamiltonians whose spectrum can be obtained
(quasi)exactly. In the following section we discuss the symmetry properties
of the general Hamiltonian (\ref{1}).

\section{Discussion: The Unified Approach}

In the previous sections transformations of the general two-mode bosonic $%
2\times 2$\ Hamiltonian were treated and its differential realization in the
Bargmann-Fock space have been obtained. In addition, we note that our
approach unifies various Lie algebraic methods. The algebraic approach to
the finite dimensional part of spectrum consists in expressing the
Hamiltonian, in terms of the generators of an Lie algebra. When the
Hamiltonian can be written in terms of the Casimir invariants of the
algebraic structure then the eigenvalue problem $H\psi =E\psi $ can be
solved in the closed form, giving rise to energy formulas\cite{alhass,Wybo}.
Otherwise, in general, the spectrum of $H$ cannot be calculates in the
closed form. If it is written in terms of the bilinear combinations of the
generators of the Lie algebra then the eigenvalue problem $H\psi =E\psi $
must be solved numerically or one can obtain quasi exact solutions.

A convenient way to construct a spectrum generating algebra is to introduce
the boson representations of the corresponding Lie algebra. Let us turn our
attention to the purely bosonic part of the Hamiltonian (\ref{1}). This can
be obtained by setting $\beta =\gamma =\delta =0$. Under some certain
conditions one can obtain and Hamiltonian which include bilinear products of
the bosonic operators:

\begin{equation}
a_{1}^{+}a_{1},a_{2}^{+}a_{2},a_{1}^{+}a_{2},a_{2}^{+}a_{1}  \label{34}
\end{equation}%
generates the Lie algebra $su(2)$ and one can recast these four generators
in a more familiar form by introducing the three generators%
\begin{equation}
J_{0}=\frac{1}{2}\left( a_{1}^{+}a_{1}-a_{2}^{+}a_{2}\right)
,J_{+}=a_{1}^{+}a_{2},J_{-}=a_{2}^{+}a_{1}.  \label{35}
\end{equation}%
The one variable differential realizations of the generators (\ref{35}) can
be obtained by the transformation procedure given in the previous sections
and they play an important role to the QES of the Hamiltonian which
underlying $su(2)$ symmetry.

The other important Lie algebra that included by the Hamiltonian (\ref{1})
is the $su(1,1)$ algebra. Bosonic representations of the $su(1,1)$ algebra
is given by%
\begin{equation}
K_{0}=\frac{1}{2}\left( a_{1}^{+}a_{1}+a_{2}^{+}a_{2}+1\right)
,K_{+}=a_{1}^{+}a_{2}^{+},K_{-}=a_{2}a_{1}.  \label{36}
\end{equation}%
Therefore by the appropriate choice of the parameter $\alpha _{i,j,k,l}$ of (%
\ref{1}) one can construct the Hamiltonian possesses the symmetry of the $%
su(1,1)$. The transformations of the operators of the $su(1,1)$ lead to the
one dimensional differential realizations of the algebra. Both in the $su(2)$
and $su(1,1)$ algebras the transformed operators are in the form of the
operators of the $sl$-algebra. One can treat bound state problems by using
the $su(2)$ algebra and scattering state problems $su(1,1)$ algebra. Thus
one can have transition from one bound state to another and from a
scattering state to another. In order to calculate from bound states to
scattering states we need a larger algebra\cite{alhass,alhas2,alhas3}. We
can construct this algebra by considering the bilinear combinations of the
bosonic operators%
\begin{equation}
a_{1}^{+}a_{1},a_{2}^{+}a_{2},a_{1}^{+}a_{2},a_{2}^{+}a_{1},a_{1}^{+}a_{2}^{+},a_{2}a_{1},a_{1}a_{1},a_{2}a_{2},a_{1}^{+}a_{1}^{+},a_{2}^{+}a_{2}^{+}.
\label{37}
\end{equation}%
One can show that these $10$ operators close under the sympletic algebra $%
Sp(4,R)$. The algebra $Sp(4,R)$ contains both bound state algebra $su(2)$
and scattering state algebra $su(1,1)$. Therefore the algebra provides a
unified treatment within both bound and scattering states. Its generators
can connect all states in the same potential. It thus provide a unified
approach to the one dimensional problems.

In addition to the Lie algebras $su(2),su(1,1)$ and $Sp(4,R)$ the
Hamiltonian (\ref{1}) include two other Lie algebras; $osp(2,1)$ and $%
osp(2,2)$\cite{chen1,chen2}. The natural step to relate the Hamiltonian (\ref%
{1}) and $osp(2,1)$ algebra is to express the Hamiltonian as linear and/or
bilinear combinations of the operators of the $osp(2,1)$ algebra. The
algebra $su(2)$ can be extended to the $osp(2,1)$ with the operators: 
\begin{equation}
V_{+}=\sigma _{+}a_{2},V_{-}=-\sigma _{+}a_{1},W_{+}=\sigma
_{-}a_{1}^{+},W_{-}=\sigma _{-}a_{2}^{+}  \label{38}
\end{equation}%
or by introducing the operators 
\begin{equation}
V_{+}=\sigma _{-}a_{2},V_{-}=-\sigma _{-}a_{1},W_{+}=\sigma
_{+}a_{1}^{+},W_{-}=\sigma _{+}a_{2}^{+}  \label{39}
\end{equation}%
The transformation of the operators of the $osp(2,1)$ by the operators $S$
or $T$ gives its one variable $2\times 2$ matrix realizations which are
useful for practical applications.

One of the major symmetry group candidates for spin one-half particles is
the supergroup $osp(2,2)$ which has four even and four odd generators. Its
even generators can be represented by bosons while odd generators are
represented by combinations of the fermions and bosons. The superalgebra $%
osp(2,2)$ might be constructed by extending $su(1,1)$ algebra with the
fermionic generators. It is possible to express two set of fermionic
generators to extend the $su(1,1)$ algebra to the $osp(2,2)$ algebra. These
are given by 
\begin{eqnarray}
V_{+} &=&\sigma _{-}a_{2}^{+},V_{-}=\sigma _{-}a_{1},W_{+}=\sigma
_{+}a_{1}^{+},W_{-}=\sigma _{+}a_{2}  \label{40a} \\
V_{+} &=&\sigma _{+}a_{2}^{+},V_{-}=\sigma _{+}a_{1},W_{+}=\sigma
_{-}a_{1}^{+},W_{-}=\sigma _{-}a_{2}.  \label{40b}
\end{eqnarray}

Among these let us also mention here, the procedure given here can also
related to the nonlinear Lie algebras that have been great deal of interest
because of their several significant applications in several branches of
physics. Let us illustrate these relation on an example. Consider The
effective Hamiltonian, which represents the Jaynes-Cummings model with Kerr
nonlinearity, have been expressed as\cite{buzek} 
\begin{equation}
H=\omega a^{+}a+\frac{1}{2}\omega _{0}\sigma _{0}+\kappa (a^{+}\sigma
_{-}+a\sigma _{+})+\lambda a^{+}aa^{+}a  \label{ed6}
\end{equation}%
where $\kappa $ and $\lambda $ are coupling constants of the field and atom
and coupling constant of the field and Kerr medium respectively. The
Hamiltonian (\ref{ed6}) can be expressed in terms of the Hamiltonian (\ref{1}%
) by an appropriate choice of parameters and the number operator of this
structure becomes:%
\begin{equation}
N=a_{1}^{+}a_{1}+\frac{1}{2}\sigma _{0}.  \label{ed7}
\end{equation}%
The invariance algebra of the Hamiltonian (\ref{ed6}) is generated by
introducing the generators,%
\begin{equation}
Y_{+}=a_{1}\sigma _{+},\quad Y_{-}=a_{1}^{+}\sigma _{-},\quad
Y_{0}=a_{1}^{+}a_{1}+\sigma _{0}  \label{ed8}
\end{equation}%
yields the commutation relation:%
\begin{equation}
\left[ Y_{0},Y_{\pm }\right] =\pm Y_{\pm },\quad \left[ Y_{+},Y_{-}\right]
=(1+2Y_{0})(Y_{0}-N)-\frac{1}{2}  \label{ed9}
\end{equation}%
The commutation relation $\left[ Y_{+},\quad Y_{-}\right] $ is a polynomial
in $Y_{0}$. Since the deformation is quadratic in $Y_{0}$, we have a
quadratic algebra. The algebras of type (\ref{ed9}) have been considered as
deformed $su(2)$ algebra. We can easily express the Hamiltonian (\ref{ed6})
in terms of the generators of the deformed $su(2)$ algebra,%
\begin{equation}
H=\omega (2N-Y_{0})+\omega _{0}(Y_{0}-N)+\kappa (Y_{+}+Y_{-})+\lambda
(2N-Y_{0})^{2}  \label{ed10}
\end{equation}
Note that the number operator $N$ is associated with the conserved quantity
of the physical system and it commutes with the generators $Y_{+}, Y_{-}$
and $Y_{0}$. Consequently, in this article, we have shown various Lie
algebraic approaches can be treated in a unified framework.

\section{Conclusion}

In this paper we have prepared a general method to obtain the solution of
two boson and one fermion Hamiltonian. By using either solution of number
operator or similarity transformation, we have been able to provide a QES of
the various physical Hamiltonians. Furthermore, it has been given that two
boson Hamiltonian can be reduced to single variable differential equation in
the Bargmann-Fock space. It is also important to mention here that the
methods given here can be used to solve higher order differential equations.

The algebras can be realized in several ways. For practical applications,
the realizations are given in terms of the boson creation and annihilation
operators. In this paper we have discussed the connection between two-boson
and one variable differential realizations of the various Lie algebras. The
realization in terms of the one variable differential equation directly
leads to the usual Schr\"{o}dinger formulation. The method given here, can
easily be extended to solve the Hamiltonians that include multi-boson or
multi fermion-boson systems. We have presented a first step toward the
extension of the formulation to obtain solution of the various physical
problems. Finally, we have shown that the technique given here provides a
unified approach to the various algebraic techniques.

\textbf{Acknowledgements}

The authors would like to thank the referee for very constructive comments

and suggestions.


\begin{thebibliography}{99}
\bibitem{klim2} Klimov A B and Sanchez-Soto L L (2000) Pyhs. Rev. A 61 063802

\bibitem{kara4} Karassiov V P, Gusey A A and Vinitsky S I (2001)
Hep-quant/0105152

\bibitem{gab} Gabriel Alvarez and Ramon F Alvarez-Estrada (2001) J. Phys. A:
Math. Gen. 34 10045

\bibitem{alv} Alvarez G and Alvarez-Estrada R F (1995) J. Phys. A: Math.
Gen. 28 5767

\bibitem{beck} Beckers J, Brihaye Y and Debergh N (1999) J. Phys. A: Math.
Gen. 32 2791

\bibitem{brihaye} Brihaye Y, Kosinski P (1994) J. Math. Phys. \textbf{35}
3089

\bibitem{tjin} Tjin T (1992) Int. J. Mod. Phys. A7 6175

\bibitem{abd} Abdesselam B, Beckers J, Chakrabarti A and Debergh N (1996) J.
Phys. A 29 3075

\bibitem{sunil} Sunilkumar V, Bambah B A, Jagannathan R, Panigrahi P K and
Srinivasan (2000) J. Opt. B: Quantum Semiclass. Opt. 2 126

\bibitem{turb1} Turbiner A V and Ushveridze A G (1987) Phys. Lett. A 126 181

\bibitem{dolya1} Dolya S N and Zaslavskii O B (2000) J. Phys. A: Math. Gen.
33 L369

\bibitem{dolya2} Dolya S N and Zaslavskii O B (2001) J. Phys. A: Math. Gen.
34 5955

\bibitem{elber} Eberly J H, Narozhny N B and Sanchez-Mondragon J J (1980)
Phys. Rev. Lett. 44 1329

\bibitem{kara1} Karassiov V P (1994) J. Phys. A 27 153

\bibitem{kara2} Karassiov V P and Klimov A B (1994) Phys. Lett. A 189 43

\bibitem{delgado} Delgado J, Luis A, S\'{a}nchez-Soto L L and Klimov A B
(2000) J. Opt. B: Quantum Semiclass. Opt. 2 33 5.

\bibitem{klim} Klimov A B, Romero J L, Delgado J and S\'{a}nchez-Soto L L
(2003) J. Opt. B: Quantum Semiclass. Opt. 5 34 6.

\bibitem{kara3} Karassiov V P (1998) Phys. Lett. A 238 19

\bibitem{band} Bandilla A, Drobny G and Jex I (1996) Phys. Rev. A 53 507

\bibitem{jurco} Jurco B (1989) J. Math. Phys. 30 1289

\bibitem{kumar} Kumar V S, Bambah B A and Jagannathan R (2001) J. Phys. A:
Math. Gen 34 8583

\bibitem{per} Perina J 1991 Quantum Statistics of Linear and Nonlinear
Optical Phenomena (Dordrecht: Kluwer) ch 10

\bibitem{baj} Bajer J and Miranowicz A (2000) J. Opt. B: Quantum Semiclass.
Opt. 2 L10

\bibitem{qu} Qu Fa, Wei Bao-Hua, Yu K W and Lui Cui-Hong 1996 J. Phys.
Condens. Matter 8 2957

\bibitem{bo} Jing-Bo ZU and Xu-Bo ZOU (2001) Chin. Phys. Lett. 18 51

\bibitem{reik} Reik H G, St\"{u}lze M E amd Doucha M (1987) J. Phys. A:
Math. Gen 20 6327

\bibitem{koc} Ko\c{c} R, T\"{u}t\"{u}nc\"{u}ler H, Koca M and K\"{o}rc\"{u}k
E (2003) Prog. Theor. Phys. 110 399

\bibitem{alhass} Alhassid Y, G\"{u}rsey F and Iachello F, (1983) Ann. Phys.
148 346

\bibitem{Wybo} Wybourne B G (1974) `` Classical Groups for Physicist`` ,
John Wiley \& Sons New York

\bibitem{alhas2} Alhassid Y and Levine R D 1978 Phys. Rev. A 18 89

\bibitem{alhas3} Alhassid Y and Koonin S E 1981 Phys. Rev. C 23 1590

\bibitem{chen1} Chen Yong-Qing (2000) J. Phys. A: Math. Gen. \textbf{33}
8071; (2001) Int. J. Theor. Phys. \textbf{40} 1113; (2000) Int. J. Theor.
Phys. \textbf{39} 2523

\bibitem{chen2} Chen Yong-Qing, Xiao-Hui Liu and Xing-Chang Song (1994)
Commun. Theor. Phys. 22 123

\bibitem{buzek} Buzek V and Jex I (1990) Opt. Cummun. 78 425
\end{thebibliography}
\end{document}